# Defect Engineering of the Water-Carbon Interface: Insights from First-Principles Calculations

Peikun Zhang, Hongbo Zhang, Ying Xu*, and Zhuhua Zhang*

State Key Laboratory of Mechanics and Control for Mechanical Structures, Key Laboratory for Intelligent Nano Materials and Devices of the Ministry of Education, and Institute for Frontier Science, Nanjing University of Aeronautics and Astronautics, Nanjing 210016, China.

* Corresponding author (email: yingxu@nuaa.edu.cn ; chuwazhang@nuaa.edu.cn )

**Abstract**

Interactions at solid-liquid interfaces are fundamental to hydrovoltaic and catalytic technologies. While carbon nanomaterials have been extensively studied, the role of abundant defects in amorphous carbon remains poorly understood. Here, using first-principles calculations, we systematically investigate how various defects (topological, point, line, edge, and doping defects) modulate water-carbon interactions. We find that armchair-edge defects with dangling bonds reduce the energy barrier for water dissociation by up to 30 times compared with pristine carbon, while di-vacancies exhibit a near-zero Gibbs free energy for hydrogen evolution, enabling potential hydrogen evolution reaction activity on a purely carbon-based interface. Moreover, mono-vacancy defects, characterized by reduced electronic localization, enhance charge transfer to evaporating water molecules by up to 12.4 times, a key factor in evaporation-induced electricity generation. Crucially, we demonstrate that coupling distinct defects produces synergistic effects that outperform individual defects, revealing a promising strategy for designing multifunctional interfaces. These insights establish a defect-engineering framework for the rational design of high-efficiency hydrovoltaic and catalytic devices.

## Introduction

The behavior of water molecules at solid surfaces underpins both the fundamental understanding of solid-liquid interfaces and a wide range of industrial applications, including catalysis and energy storage [1-3]. Among various solid-liquid interface systems, the water-carbon interface has attracted particular attention due to its structural simplicity and distinctive physical properties, such as an extremely low dielectric constant [4] and superlubricity [5]. Recently, a novel phenomenon known as the hydrovoltaic effect has emerged at this interface, enabling the direct conversion of the latent heat of water into electricity [6-8]. This effect can be driven by various forms of water motion, including droplets [9], waves [10], evaporation [11], and moisture [12], all of which are governed by the interactions between water molecules and the material surface. Graphene-based carbon nanosheets, as model systems, have been widely employed to investigate these interactions in hydrovoltaic experiments [9, 13-15]. However, their limited charge transfer efficiency at the interface hinders further development of hydrovoltaic technologies [16]. Therefore, modulating the structure of carbon substrates to achieve functionalities conducive to the hydrovoltaic effect has become a critical pursuit. Defect engineering offers a promising route, yet a systematic understanding of how specific defects modulate water-carbon interactions remains lacking.

The recent successful synthesis and in-situ atomic-scale characterization of monolayer amorphous carbon mark a new opportunity for further exploration of the water-carbon interface [17-20]. Unlike conventional carbon materials, monolayer amorphous carbon completely lacks long-range periodicity [19, 21]. Its structure features predominantly threefold coordination with broad distributions of bond lengths and bond angles, as well as abundant topological defects, such as pentagons, heptagons, and their combinations like Stone-Wales defects [17, 22-25]. These structural characteristics endow it with unique and more reactive physical properties, which are expected to lead to distinct adsorption and transport behaviors of water molecules compared to traditional graphene or graphene oxide surfaces [26-29]. This positions monolayer amorphous carbon as a promising platform for hydrovoltaic applications. Notably, defects in graphene have been shown to play a critical role in modulating water-related reactions, including oxygen evolution and hydrogen evolution [30-31]. However, the adsorption mechanism governing water molecule behavior on monolayer amorphous carbon surface remains unclear, presenting a key challenge for its application in the hydrovoltaic effect.

In this work, we deconstructed monolayer amorphous carbon into a set of representative defects and systematically investigated their interactions with water molecules using first-principles calculations. By examining the adsorption configurations and electronic properties at various defect sites, we elucidate how different defects (topological, point, line, edge, and doping defects) modulate water adsorption, charge transfer, and subsequent reaction pathways. Our results reveal that most water molecules act as electron acceptors when interacting with defective graphene, whereas they behave as an electron donor on N-doped graphene, corresponding to distinct adsorption orientation. For mono-vacancy, zigzag-edge and armchair-edge defects, the water dissociation energy barriers are below 0.6 eV, much lower than that of pristine graphene (3.6 eV). During water molecule desorption from the graphene surface, defective graphene can increase the amount of charge carried away by departing water molecules. In particular, the mono-vacancy enables a single water molecule to carry away an order of magnitude more charge than on pristine graphene, which is beneficial for evaporation-induced electricity generation. Our work paves the way for enhancing the hydrovoltaic effect by modulating the structure of monolayer amorphous carbon through defect engineering.

## Methods

The calculations were performed within the framework of density functional theory (DFT) as implemented in the Vienna Ab-initio Simulation Package (VASP) code by using the projector augmented wave method with the Perdew-Burke-Ernzerhof (PBE) exchange-correlation functional [32-35]. The energy cutoff of the plane-wave expansion was set to be 500 eV. The van der Waals (vdW) interactions between the water molecule and the graphene were described by the D3 method [36]. The D2 and optB86 methods were also employed to check the results, and no qualitative change was found [37-38]. The atomic positions and geometries were fully relaxed with the conjugate gradient method until the residual force on each atom dropped below the threshold value of 0.01 eV/Å and the energy convergence was within $10^{-5}$ eV. A vacuum space of 15 Å was set to isolate neighboring periodic images. The energy barriers of water dissociation were calculated using the climbing-image nudged elastic band (CI-NEB) method [39-40].

For pristine and defective graphene with periodic structures (hexagon, Stone-Wales, mono-vacancy, di-vacancy, ad-dimer, B-doped graphene, and N-doped graphene), an 8×8 supercell with a 3×3×1 k-point grid was employed. For pentagon and heptagon defects, which require non-planar geometries, cone and pringle models containing over 100 carbon atoms were constructed, respectively, with edges passivated by hydrogen. Due to their non-periodic nature, a 1×1×1 k-point grid was employed. For the 5|7 dislocation, a twin boundary was constructed within a 9×5 supercell, and a 3×7×1 k-point grid was adopted. For zigzag-edge and armchair-edge structures, periodicity was defined along the *x*-direction. Each unit cell contained eight hexagonal rings, while the nanoribbons had a width of five hexagonal rings, with a k-point grid of 6×1×1 used. For magnetic systems (mono-vacancy and zigzag-edge, Figure S1), spin-polarized calculations were employed, while for other systems, non-spin-polarized calculations were used. We performed convergence tests on the k-point sampling, cutoff energy, and vacuum spacing to ensure the reliability of our results (Figure S2).

The adsorption energy is defined as:

$$E_{\text{ads}} = E_{\text{total}} - E_{\text{graphene}} - E_{\text{H2O}}$$

Where $E_{\text{total}}$ is the total energy of the adsorption system, $E_{\text{graphene}}$ and $E_{\text{H2O}}$ are the energies of the graphene and an isolated water molecule, respectively.

The definition formula of charge density difference is:

$$\Delta\rho = \rho_{\text{total}} - \rho_{\text{graphene}} - \rho_{\text{H2O}}$$

Where $\rho_{\text{total}}$ is the total charge density of the adsorption system, $\rho_{\text{graphene}}$ and $\rho_{\text{H2O}}$ are the charge densities of the graphene and an isolated water molecule, respectively.

The charge transfer $\Delta q$ is defined as the net number of electrons gained by the water molecule after adsorption:

$$\Delta q = q_{\text{H2O-G}} - q_{\text{H2O}}$$

Where $q_{\text{H2O-G}}$ is the number of electrons carried by the water molecule after adsorption, and $q_{\text{H2O}}$ is the number of electrons of the isolated water molecule. A positive $\Delta q$ means that the water molecule gains electrons and becomes negatively charged.

The hydrogen adsorption energy is defined as:

$$\Delta E_{H} = E_{H^{*}graphene} - E_{graphene} - 1/2E_{H2}$$

Where $E_{H^{*}graphene}$ is the energy of graphene with an adsorbed H atom, $E_{graphene}$ is the energy of defective graphene, and $E_{H2}$ is the energy of a gas-phase $H_2$ molecule.

The Gibbs free energy ($\Delta G_{H}$) is defined as:

$$\Delta G_{H} = \Delta E_{H} + \Delta E_{ZPE} - T\Delta S_{H}$$

Where $\Delta E_{H}$ is the hydrogen adsorption energy, $\Delta E_{ZPE}$ is the zero-point energy difference, $T$ is the temperature, and $\Delta S_{H}$ is the entropy difference for hydrogen between the adsorbed state and the gas phase.

The crystal orbital Hamilton populations (COHP) analysis was implemented in the LOBSTER software [41], which takes the delocalized planewaves as the input. One important quantity accessible in this method is the integrated value of COHP (ICOHP), which can be employed to approximately quantify the bond strength. For clarity, both quantities above are presented as their negative values. Our study focuses on the COHP between the O/H atoms of the water molecule and the neighboring substrate atoms, including C-O/C-H interactions for pristine and carbon-defect systems and the corresponding interactions involving C, B, or N atoms for the B- and N-doped systems. Distance cutoffs slightly larger than the sum of the corresponding vdW radii were used to include the relevant adsorbate-substrate interactions (4 Å for C-O and 3 Å for C-H). For each system, the COHP contributions from the included atom pairs were averaged.

The time-dependent density functional theory (TD-DFT) calculations were performed using the DFTB+ software [42-43]. Interatomic Coulomb interactions were treated with the DFTB3 method in conjunction with the 3ob parameter sets [44]. The water molecule was pulled away from the graphene surface at a velocity of 0.1 Å/fs, with a total simulation time of 50 fs and a time step of 0.005 fs. At each electronic dynamics step, the partial charges were iteratively updated until a predefined convergence criterion was achieved. Atomic partial charges were evaluated using Mulliken population analysis at each time step. This approach may introduce deviations in describing charge transfer on defective carbon, and therefore we treat the results as a semi-quantitative trend analysis.

**Results and discussion**

Figure 1a displays the monolayer amorphous carbon synthesized by Deng et al. on a Si/$SiO_2$ substrate, exhibiting structural disorder and electrical insulation [19]. To capture the structural characteristics, we constructed a large-scale atomic model of monolayer amorphous carbon through Monte Carlo (MC) simulations, incorporating conventional hexagons (HEX), as well as carbon pentagons, heptagons, their combined defect structures, vacancies, and adatoms (Figure 1b). To study the impact of individual defects, we created separate models and classified them into four categories: topological, point, line, and edge defects (Figure 1c). Topological defects include carbon pentagons (PNT) and heptagons (HPT). Point defects comprise Stone-Wales defects (SW), mono-vacancies (MV), di-vacancies (DV), and ad-dimers (AD), while the line defect corresponds to the 5|7 dislocation (5|7). Edge defects include zigzag (ZZ) and armchair (AC) edges. Additionally, we included two common doping defects which have been experimentally realized in amorphous graphene or carbon materials [20, 45], B-doped graphene (B-g) and N-doped graphene (N-g), to provide a broad perspective on water-carbon interactions. These defective graphene systems exhibit distinct electrical properties. For example, the SW defect behaves as a semiconductor with a band gap of 0.17 eV (Figure S3), while the electron localization function (ELF) at the DV defect is significantly reduced (Figure S4). These differences suggest that the interactions between water molecules and various defects are also likely to be distinct.

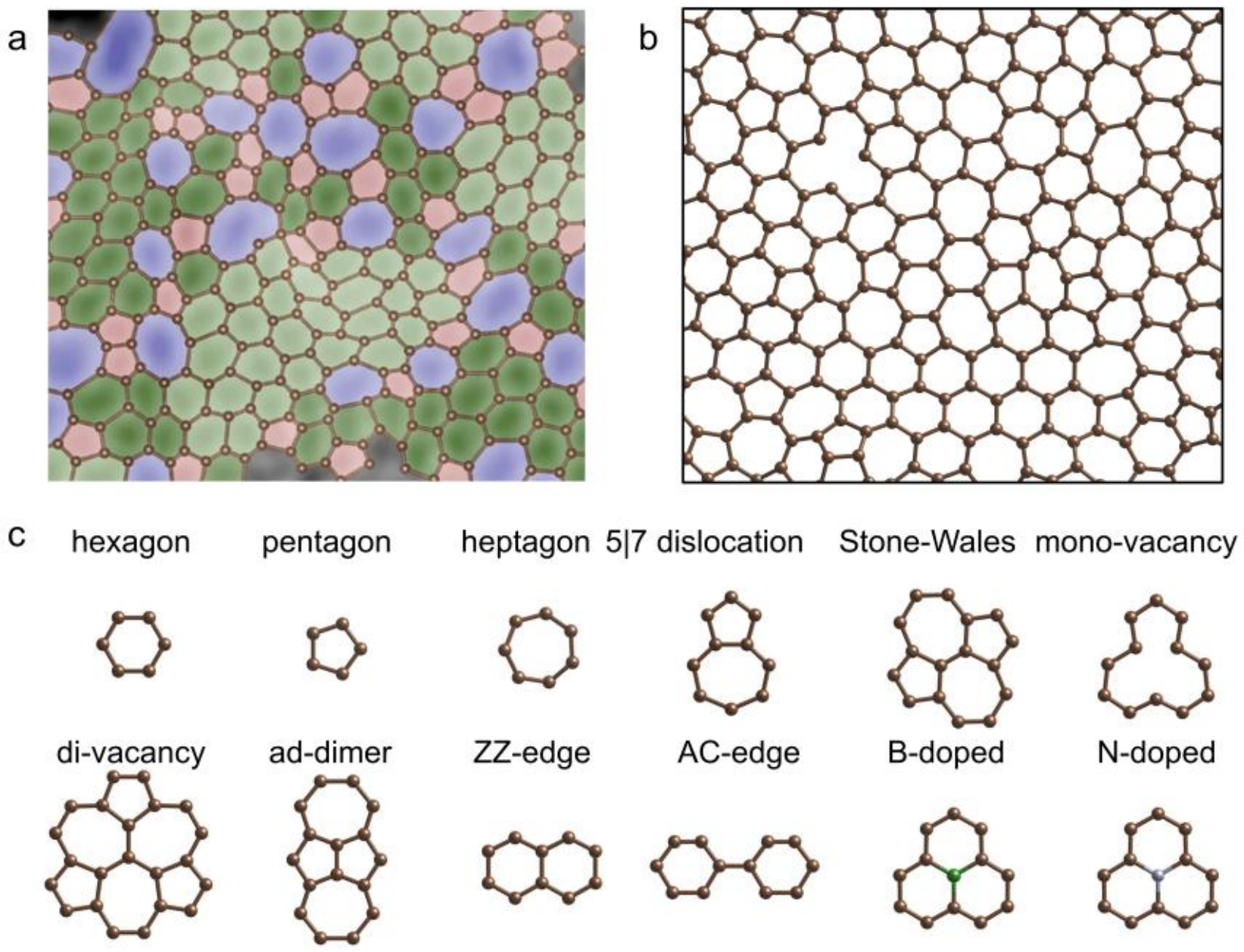

**Figure 1** Structure and defects of monolayer amorphous carbon. (a) Color overlay on the STEM image of monolayer amorphous carbon. Copyright 2025 Springer Nature. (b) Monolayer amorphous carbon structure obtained from MC simulations. (c) Atomic configurations of various defects in monolayer amorphous carbon.

Considering the polarity of water molecules, we examined three possible orientations for a water molecule adsorbed on graphene surface, denoted as 0-leg, 1-leg, and 2-leg configurations (Figure 2a). The 0-leg, 1-leg, and 2-leg configurations have zero, one, or both hydrogen atoms pointing toward the graphene surface, respectively. To ensure computational reliability, we performed benchmark calculations of water adsorption on an ideal graphene lattice using different van der Waals corrections. For comparison, we also considered reference data from other studies using the diffusion Monte Carlo (DMC) method and random phase approximation with GW single excitations (RPA+GWSE) [28]. The calculated adsorption heights and energies are shown in Figure S5, revealing consistent trends across all methods. Considering both accuracy and computational cost, we selected DFT-D3 as our vdW correction method, which yields results in good agreement with previous studies [26, 28]. Due to symmetry reduction induced by defects, numerous inequivalent adsorption sites exist (Figure S6). We performed a structural relaxation scan across all sites to identify the most stable adsorption sites and configurations for the three adsorption states. The corresponding adsorption heights and energies are presented in Figures 2b-c, with the optimized adsorption structures shown in Figure 2d. For pristine graphene, the adsorption energy of a water molecule is -0.16 eV. Defects such as MV and B-g exhibit stronger interactions, with the MV defect reaching -0.23 eV. Notably, since real amorphous graphene possesses long-range disorder and local strain, we performed water adsorption calculations on the optimized structure previously generated by MC simulation. The results show that the adsorption energies obtained from the isolated-defect models show the same trend as those for the corresponding local environments in the disordered model, with a difference of less than 15% in absolute relative error (Figure S7).

Regarding the adsorption state, a water molecule prefers the 2-leg adsorption configuration on pristine graphene and most defective graphene structures, including PNT, HPT, 5|7, SW, MV, DV, AD, and B-g. This preference arises because the positively charged H atoms face the graphene surface, which carries a negatively charged π-electron cloud. Such an orientation enables attractive electrostatic interaction between opposite charges, thereby lowering the system's energy. In addition, it maximizes the contact area between the water molecule and graphene, further enhancing vdW interactions. In contrast, for the two edge configurations, ZZ and AC, the water molecule adopts the 1-leg configuration. This difference stems from the distinct nature of the graphene edges compared to the basal plane. Carbon atoms at the edges possess dangling bonds due to unsaturation, making them chemically highly active and prone to capture H atoms for self-passivation. Consequently, one H atom is oriented towards the edge C atom. For N-g, the optimal configuration is the 0-leg geometry, in which the O atom points toward graphene. This behavior is attributed to the high electronegativity of the N atom, which withdraws electrons from the surrounding C atoms, leaving the adjacent graphene region partially positively charged and thus attracting the negatively charged O atom. Furthermore, for the same water molecule orientation configuration, taking 2-leg as an example, an approximately linear relationship exists between adsorption energy and adsorption height: a lower adsorption height corresponds to a larger adsorption energy, consistent with the characteristics of vdW interactions (Figure S8).

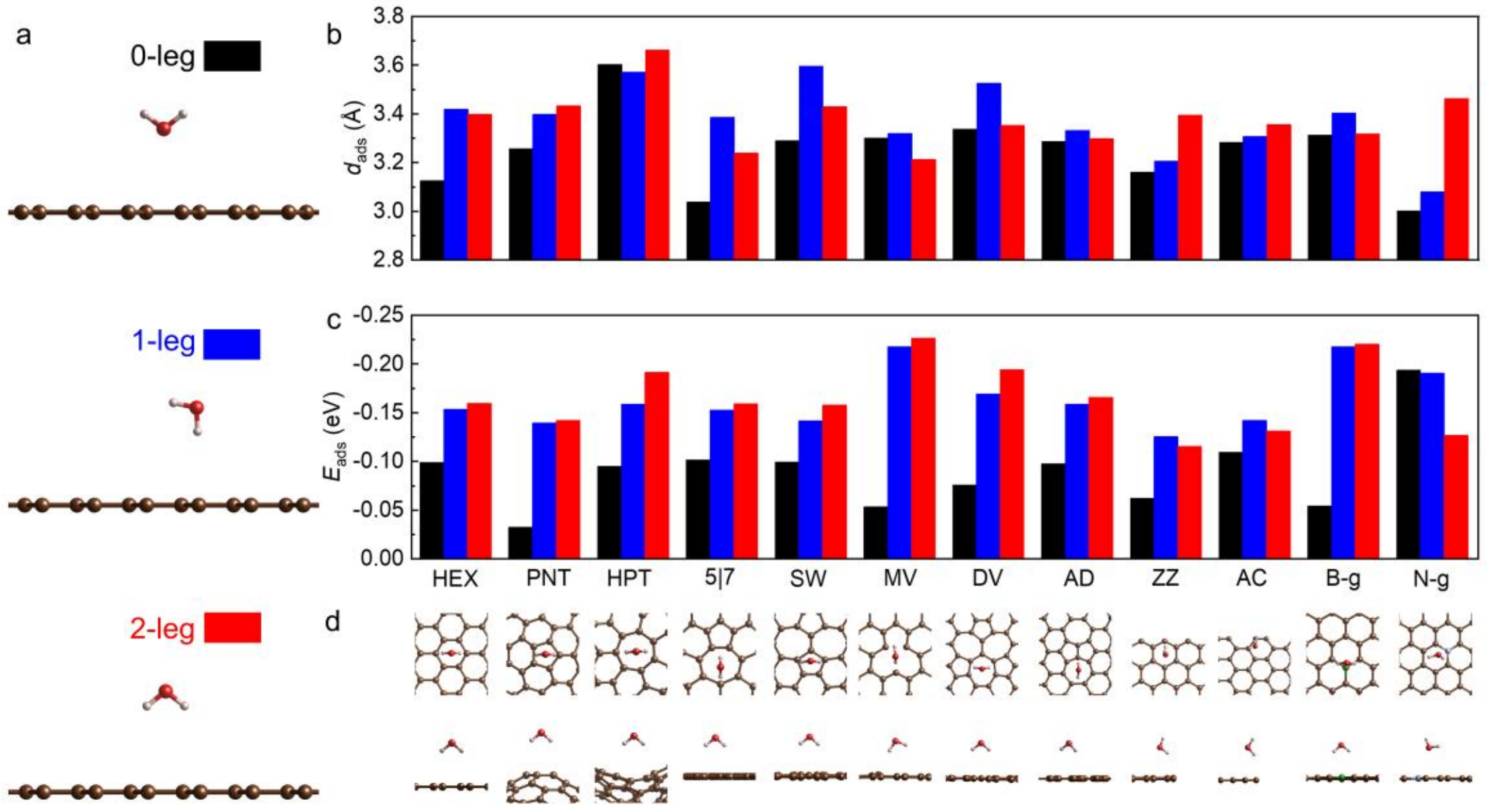

**Figure 2** The adsorption status of a water molecule on graphene. (a) Side views of three water adsorption orientations (0-leg, 1-leg, and 2-leg) on graphene. (b) Height between the adsorbed water molecule and graphene ($d_{ads}$) and (c) adsorption energy ($E_{ads}$) of the three orientations on graphene with various defects. (d) Top and side views of the optimal configuration for each defect structure.

To gain deeper insight into the interaction between a water molecule and various graphene defects, we analyzed the charge transfer between the optimal water configuration and the graphene substrate (Figure 3a). For the 2-leg orientation, an electron depletion region commonly appears on the lower surface of graphene, i.e., the surface not in contact with water. The depleted electrons accumulate substantially in the region between the water molecule and graphene, indicating that negative charge governs the interaction between them. Regarding the water molecule itself, the H atoms lose a small amount of charge while the O atom gains a small amount, enhancing the electrostatic interaction. Notably, due to their unique three-dimensional structural characteristics, the charge distribution on the lower surface of defects containing PNT and HPT differs from others, yet the overall trend remains consistent with the above description. For the 1-leg orientation at the graphene edge, the charge transfer leads to a distinct in-plane charge distribution. Specifically, at the interface where the downward-pointing H atom contacts the graphene edge, the H atom loses electrons and the graphene edge gains electrons. On the opposite side, the O atom acquires electrons from graphene, with negative charge dominating the interaction between the graphene edge and the water molecule. Unlike the 2-leg case, the lower surface of graphene exhibits a weak electron gain region. For the 0-leg orientation on N-g, the O atom gains a substantial number of electrons from graphene, giving rise to a very pronounced charge depletion region between the water molecule and graphene. Moreover, electron gain on the lower surface of graphene is more pronounced compared to the edge case. Overall, the charge transfer characteristics differ distinctly among the three water molecule configurations interacting with graphene.

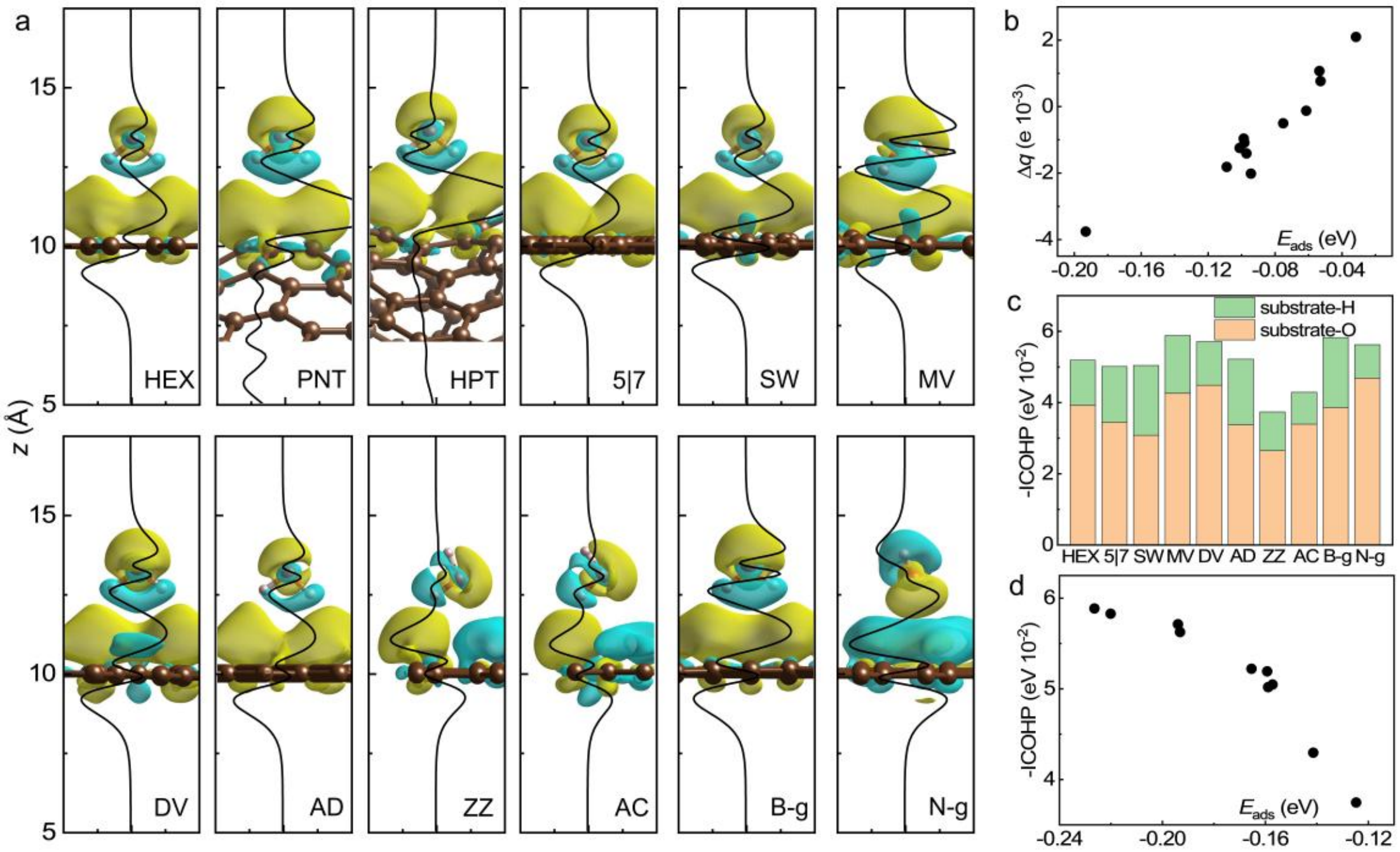

**Figure 3** Charge transfer and bonding analysis. (a) Isosurface plot (0.001 e Å$^{-3}$) of the calculated charge density difference of the graphene induced by water adsorption, along with the corresponding plane-averaged profiles. Charge accumulation and depletion regions are distinguished by yellow and blue, respectively. (b) Calculated charge transfer between the water molecule and graphene as a function of adsorption energy $E_{\text{ads}}$ for the 0-leg configurations. (c) Integrated -COHP (-ICOHP) for each defect structure. (d) The -ICOHP values are plotted as a function of the adsorption energy.

Bader charge analysis reveals a linear relationship between adsorption energy and charge transfer for a given water orientation. For the 0-leg configuration, charge transfer correlates positively with adsorption energy (Figure 3b), indicating that greater charge transfer corresponds to weaker adsorption, whereas for the 1-leg and 2-leg configurations, it correlates negatively (Figure S9). To further elucidate the orbital hybridization, we calculated the density of states (DOS, Figure S10) and projected density of states (PDOS, Figure S11). The water molecule shows clear hybridization with the substrate at 1.5-4.0 eV below the Fermi level, mainly from the O 2p orbital and the C 2p orbital (Figure S11).

In addition, LOBSTER was used to calculate the COHP between O/H atoms and neighboring substrate atoms. The bonding features identified from the -COHP analysis are broadly consistent with the energy regions showing strong DOS/PDOS overlap, providing complementary evidence for the electronic coupling between water and the defective graphene substrate (Figure S12). Since the adsorption interaction is predominantly noncovalent, the corresponding bonding contributions are substantially weaker than those associated with conventional covalent bonds. Integration of the -COHP curves up to the Fermi level gives the -ICOHP values, which quantify the net bonding contribution within the selected atom pairs, i.e., larger positive -ICOHP values indicate stronger bonding contributions. The calculated -ICOHP values are summarized in Figure 3c and further separated into substrate-O and substrate-H contributions. All water adsorption configurations exhibit a net bonding contribution, with the substrate-O interaction making the dominant contribution. This result is consistent with the PDOS analysis, which indicates that the O 2p states are primarily responsible for the electronic coupling between the water molecule and the substrate. More importantly, the calculated -ICOHP values show a clear correlation with the adsorption energies across the different defect sites (Figure 3d), with stronger bonding contributions corresponding to stronger water-substrate interactions, further supporting our overall analysis of the bonding strength of the systems.

Meanwhile, due to the high stability of the pristine graphene surface, water dissociation is exceptionally difficult, with an energy barrier exceeding 3 eV. However, considering that defects endow the graphene surface with higher reactivity, we explored whether water dissociation could be facilitated, potentially enabling catalytic applications. To this end, we performed transition state calculations for water dissociation on the 12 configurations mentioned above to identify which defects are more favorable for water dissociation. The corresponding dissociation pathways and energy barriers are presented in Figure 4a. Pristine graphene exhibits the highest dissociation energy barrier of 3.6 eV. Defects such as PNT and SW show a moderate reduction, with barriers lowered by approximately 0.2-1 eV. Although these defects alter the local environment, they largely preserve the $sp^2$ hybridization of the C atoms themselves, resulting in only a modest increase in surface activity. In the case of B-g, the introduction of the B atom induces a local electron deficiency in graphene, significantly enhancing surface activity and reducing the energy barrier to 2.1 eV. A similar trend is observed for the DV defect. Although the C atoms retain $sp^2$ hybridization, the removal of two carbon atoms reduces the local electron density, further lowering the energy barrier to 1.9 eV. The most pronounced reductions are observed for the three configurations with dangling bonds: MV, ZZ, and AC. Their energy barriers are all below 0.6 eV, with AC exhibiting a remarkably low barrier of 0.12 eV. This indicates that water dissociation occurs readily on these surfaces, representing a qualitative improvement over pristine graphene. This enhanced activity arises because, during the dissociation process, the breaking of an O-H bond also generates two dangling bonds, increasing the energy. The inherent dangling bonds present in these three defect configurations can effectively combine with the dangling bonds of $OH^-$ and $H^+$, forming covalent bonds while simultaneously passivating both the water molecule and the defect itself. This results in a substantial reduction of the energy barrier and yields a final state energy lower than that of the initial state. The high activity of edges and vacancies may also be applicable to other material systems that are inherently inert towards water, such as $MoS_2$.

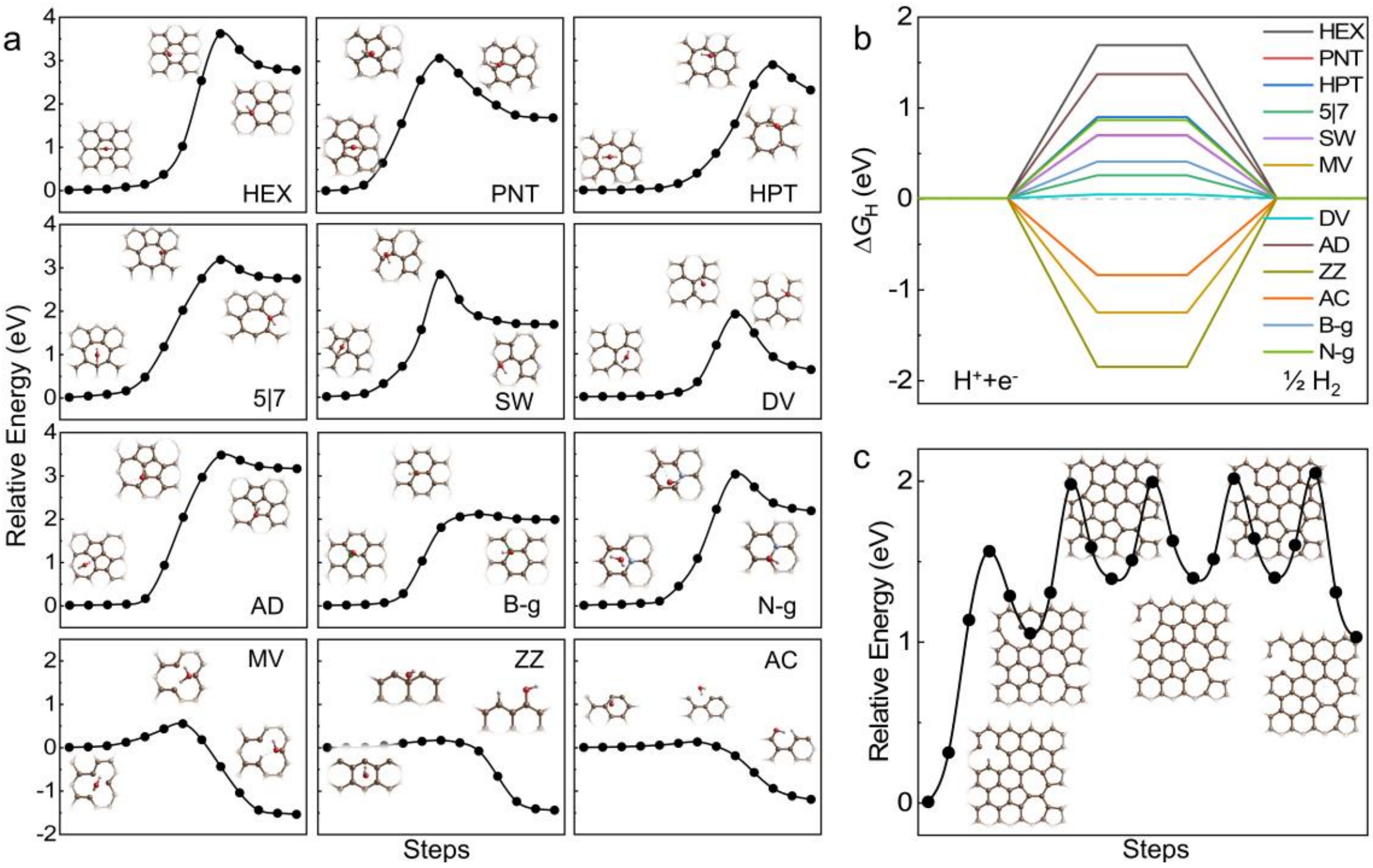

**Figure 4** Water dissociation on different defects. (a) Reaction pathways of water molecule dissociation on each defect. The insets show top views of the initial state, transition state, and final state. (b) Calculated $\Delta G_H$ for hydrogen adsorption on each defect. (c) Migration path and corresponding energy barrier for an H atom migrating from the MV to the DV site. The insets illustrate the migration process.

Water dissociation is the first step towards catalyzing the hydrogen evolution reaction (HER), followed by the reduction and desorption of hydrogen. We calculated the $\Delta G_H$ of hydrogen reduction and desorption at different sites for all defects. Given the large number of sites, we selected only the most favorable adsorption site for each defect, i.e., the one with $\Delta G_H$ closest to zero (Figure 4b). Consistent with the trend observed for water dissociation energy barriers, defects such as HPT and AD exhibit low surface activity, making hydrogen adsorption difficult. As a result, their adsorption $\Delta G_H$ values are mostly above 0.5 eV, reaching as high as 1.7 eV. For the electron-deficient B-g and DV, the absence of sufficient charge enhances the surface activity and significantly lowers the energy barrier for H adsorption. For B-g, the $\Delta G_H$ is only 0.4 eV. As the electron deficiency becomes more pronounced in DV, the $\Delta G_H$ decreases to as low as 0.05 eV, approaching the ideal $\Delta G_H$ value of Pt/C. This indicates that the reduction and evolution of $H_2$ are thermodynamically more favorable on these defect sites. For the MV, ZZ, and AC systems, all of which possess dangling bonds, surface activity is further substantially enhanced. These systems exhibit very strong hydrogen adsorption, with $\Delta G_H$ below -0.8 eV. In this case, H acts more like a passivating agent, being strongly adsorbed at the vacancy site and difficult to desorb.

Based on the calculations of water dissociation barriers and Gibbs free energies, and considering the low dissociation barrier of MV and the optimal $\Delta G_H$ of DV, we designed a bifunctional graphene containing both types of defects (Figure S13). In this system, water can dissociate with low energy consumption at the MV site, after which the H atom migrates to the DV site for reduction, enabling intrinsic HER activity in graphene. In this scenario, the transport of the H atom between the two vacancy sites becomes critically important. To evaluate this, we calculated the migration process of one H atom on the dual-defect graphene (Figure 4c). The migration path with the lowest energy barrier involves the H atom moving from a C atom top site, via a bridge site, to another C atom top site. The rate-determining step of the entire process is the first step, where the H migrates from the C atom with a dangling bond to an adjacent C atom with saturated bonds, with a barrier of 1.5 eV. Subsequent migration steps are quite similar, with energy barriers around 0.7 eV, until the H reaches the reduction site at the DV. These findings suggest that HER becomes potentially achievable even on graphene without any metal doping.

Furthermore, considering that graphene possesses excellent elasticity and can sustain large tensile strain, and that such strain may intrinsically exist in amorphous carbon due to its structural diversity, we investigated the effect of strain on adsorption. We applied biaxial strains ranging from 2% to 8% to all defect models (Figure S14). We found that, except for AD and ZZ, the adsorption energy of water on most defective structures initially increases and then decreases with increasing strain. This trend arises because the applied strain reduces the carbon atom density and electron density, thereby weakening the vdW interaction. However, beyond a certain strain threshold, the adsorption configuration of the water molecule undergoes a slight reorientation (Figure S15), enhancing electronic conjugation with the graphene surface and leading to a decrease in adsorption energy (Figure S16a). In contrast, for AD and ZZ, the water molecule maintains its original orientation even under 8% strain. Strain modulation also significantly influences the $\Delta G_H$ for HER (Figure S16b). Similarly, except for edge defects, the $\Delta G_H$ generally decreases with increasing strain. Under 2% tensile strain, the $\Delta G_H$ of the DV approaches 0, while B-g reaches a value of 0.06 eV under 8% tensile strain.

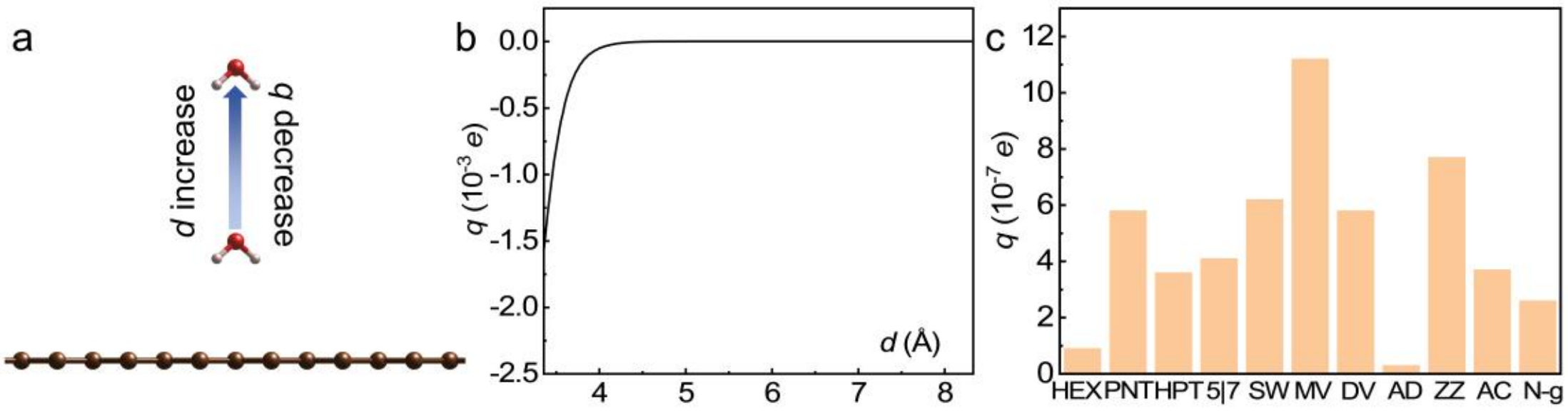

**Figure 5** Change in the charge of a water molecule during evaporation. (a) Schematic diagram of a water molecule evaporating from the graphene surface. (b) Calculated charge carried by a water molecule as a function of its distance from pristine graphene. (c) The amount of charge carried away by water molecules upon leaving various defects in graphene.

Regarding the evaporation potential in hydrovoltaics, the amount of charge carried by a water molecule during its evaporation from the graphene interface directly determines the resulting current and voltage. It should be noted that macroscopic evaporation involves the disruption of the hydrogen-bond network. However, previous studies have shown that single-molecule desorption is the principal evaporation event, particularly at the contact line of water on graphene-covered substrates [46-48]. Therefore, our model focuses on the elementary step of this process: the desorption of a single water molecule from the interface. To investigate this, we employed time-dependent density functional theory (TD-DFT) to calculate the charge carried by a water molecule as it escapes from a stable physical adsorption state on graphene surface into the vacuum (Figure 5a). The DFTB3/3ob parameter set we used has been well developed and benchmarked for systems containing C, H, and O [49-52]. We also compared the Mulliken charges calculated by DFTB+ with the Bader charges calculated by VASP for the stable adsorption configurations (Figure S17). Although the absolute values differ, as expected for these two different charge-partitioning schemes, the signs and trends of the charge transfer are consistent across the two methods, indicating that the semi-quantitative analysis is reasonably reliable. Taking pristine graphene as an example (Figure 5b), the initial charge carried by the water molecule is $-1.6\times10^{-3}$ $e$. As the water molecule moves away from the graphene surface, the carried negative charge gradually decreases, reaching 0 at a distance of 4.9 Å. Beyond this point, the molecule begins to carry a positive charge, which eventually stabilizes at $9\times10^{-8}$ $e$ at approximately 6.3 Å, representing the final charge carried away upon evaporation. Using this approach, we further evaluated the charge carried by water molecules departing from other defective graphene structures (Figure 5c), excluding B-g due to the absence of B element parameters in the well-developed 3ob potential of DFTB+. Except for the AD defect, all other defects exhibit a higher charge transfer than pristine graphene. Notably, the MV and ZZ defects show enhancement by factors of 12.4 and 8.6, respectively. These results indicate that defect-rich amorphous carbon can significantly enhance evaporation-induced charge transfer. By rationally incorporating defects such as MV and ZZ into graphene, the evaporation-induced voltage and current can be effectively maximized.

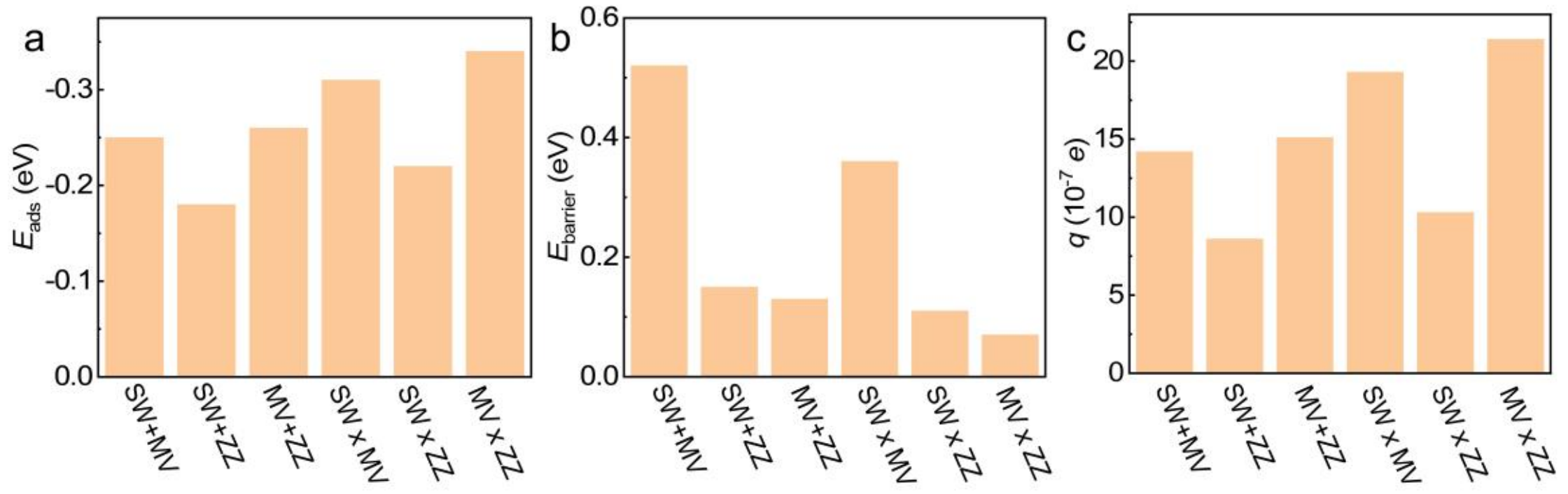

**Figure 6** Defect coupling effect. (a) Adsorption energy, (b) water dissociation energy barrier, and (c) amount of charge carried away by a water molecule for different coupled defect structures.

Defect coupling is frequently observed in amorphous carbon. To investigate this phenomenon, we constructed coupled defect structures using three representative defects (SW, MV, and ZZ) and examined their interactions with water molecules. Two types of defect coupling are considered: adjacent configurations, denoted as A+B, and superimposed configurations, denoted as A×B (A, B = SW, MV, or ZZ). The corresponding structures and the scanned adsorption sites are shown in Figure S18. The results indicate that defect coupling further strengthens the interaction with the water molecule through enhanced electron activity. Moreover, the adsorption energy of superimposed defects is generally higher than that of adjacent defects (Figure 6a). Notably, the MV×ZZ configuration achieves an adsorption energy of -0.34 eV. This enhancement is also reflected in the reduction of the water dissociation energy barrier and the increase in the charge carried away by an evaporating water molecule (Figures 6b-c). Consistently, the MV×ZZ configuration exhibits the most pronounced effect, with the water dissociation energy barrier reduced to 0.07 eV and the carried charge reaching $2.14\times10^{-6}$ $e$.

## Conclusion

We systematically investigated the interactions between water molecules and various defects in amorphous carbon using first-principles calculations. Our results reveal that most typical graphene defects enhance the water-carbon interaction compared to pristine graphene. The 2-leg adsorption configuration predominates on the basal plane, whereas the 1-leg orientation is favored at edge sites, and the 0-leg orientation is uniquely observed on N-doped graphene. Notably, a linear correlation exists between adsorption energy and charge transfer for a given adsorption orientation. Certain defects, particularly those with dangling bonds such as MV and edge structures, significantly lower the water dissociation energy barrier, with values below 0.6 eV. Moreover, DV exhibits a Gibbs free energy for hydrogen evolution approaching zero, close to the ideal $\Delta G_H$ value of Pt/C. In addition, most defects substantially increase the amount of charge carried away by a departing water molecule, with MV defects enabling an order-of-magnitude enhancement relative to pristine graphene. Furthermore, we demonstrate that coupling distinct defects, such as SW, MV, and edge structures, can produce synergistic effects that further improve adsorption strength, reduce reaction barriers, and enhance charge transfer. Our study establishes a defect-engineering framework for the rational design of carbon-based functional interfaces, offering a promising pathway toward high-efficiency hydrovoltaic and catalytic devices.

## Author Contributions

**Peikun Zhang:** Data curation; formal analysis; investigation; writing-original draft. **Hongbo Zhang:** Data curation; formal analysis. **Ying Xu:** Formal analysis; writing-review and editing. **Zhuhua Zhang:** Conceptualization; funding acquisition; supervision.

## Acknowledgement

This work was supported by the National NSF of China (12225205, 12272180, U2441272,12261160367), the Basic Research Program of Jiangsu (BK20253025, BK20243044), the National Key Research and Development Program of China (2024YFA1409600), the Research Fund of State Key Laboratory of Mechanics and Control for Aerospace Structures (Grant No. MCAS-I-0525K01), and the Special Fund for the Fundamental Research Funds for the Central Universities (NG2024026, NP2024110). We thank the support of the High-performance Computational Center at NUAA.

## Conflict of interest statement

The authors declare that they have no conflict of interest.

## Data availability statement

Supporting Information is available in the online version of the paper.